# A Case in Kenya: Unlocking bottlenecks in public health supply chains through data dashboards and enhanced governance structures

Yasmin Chandani
John Snow, Inc.
Arlington, VA, USA
yasmin_chandani@jsi.com

Elizabeth A. Bunde, PhD
John Snow, Inc.
Arlington, VA, USA
Elizabeth_bunde@jsi.com

Wambui Waithaka
John Snow, Inc.
Nairobi, Kenya
Wambui_waithaka@ke.jsi.com

Eric Wakaria
John Snow, Inc.
Nairobi, Kenya
Eric_wakaria@ke.jsi.com

James Riungu
John Snow, Inc.
Nairobi, Kenya
James_riungu@ke.jsi.com

Judith Njumwah-Kariuki
John Snow, Inc.
Nairobi, Kenya
judith_njumwah@ke.jsi.com

Alexis Strader
John Snow, Inc.
Arlington, VA, USA
Alexis_strader@jsi.com

## ABSTRACT
The link between data and governance are key to making public health supply chains more integrated and responsive in order to get life-saving commodities to those in need. In particular, considering its significant health challenges, poor maternal and child health indicators, and major recent devolution in political authority, Kenya represents a country in need of an innovative revamp of their data management and governance. John Snow, Inc. (JSI) adapted various elements of proven interventions to build a customized structure to support routine data collection in order to drive decision making around supply chain improvement.

## 1. INTRODUCTION

### 1.1 Public health supply chains, governance and data

Public health programs in the developing world that have committed to achieving Sustainable Development Goals[1] must strive to build integrated, responsive supply chains that deliver life-saving commodities to patients in need. Given resource constraints, optimizing the performance of these supply chains will require quality, available, and timely data that can be rapidly transformed into usable information for operational and strategic decisions. Supply chain and health managers, unused to having timely and comprehensive access to actionable management data, must learn how to use this data to improve health supply chain outcomes.

### 1.2 Kenya public health status and political climate

Kenya has seen significant progress in its health indicators in recent decades, but challenges remain. Child mortality remains high, with an under-five years of age mortality rate of 51 per 1,000 live births (2014) and infant mortality of 37 per 1,000 live births (2014). Maternal mortality is also high at 362 per 100,000 live births (2014)[2]. These challenges, along with the continued high burden of infectious diseases and emerging non-communicable diseases must be addressed in order to accelerate improved health outcomes for the citizens of Kenya.

In 2010, Kenya inaugurated a new constitution, which created 47 devolved units, involving significant devolution of authority, responsibility, and funding for government services to the county level. Each county is further divided into sub-counties (equivalent to districts), 285 in total. While devolution has launched a decentralized model for local decision making and management of the health system to reflect and address county-specific needs, it has also disrupted health service delivery and led to fragmentation of availability of data and of health supplies, especially for contraceptives and vaccines.

### 1.3 Addressing the data gap

Kenya lacks a logistics management information system that can serve as the heartbeat for all its supply chains. However, some logistics data is collected via DHIS2, a health information system that has been adopted as the de facto health electronic information system for the country, with the majority of countries submitting some supply chain data for contraceptives and vaccines on a monthly basis. Data in DHIS2 is often of variable quality and timeliness, hampering the ability of national and county managers to act on the data. With the devolution of the health sector, responsibility and budgetary authority has shifted to county and sub-county leaders. As a result, strengthening analytic and leadership capacity for supply chain management at the county level is key to building strong, well-performing supply chains that deliver products to health facilities.

## 2. APPROACH

John Snow, Inc. (JSI) adapted elements of proven interventions that have been successfully implemented elsewhere and that have resulted in significant reductions in stockout rates or improvements in product availability – both key outcome metrics for measuring supply chain system performance[3,4].

JSI introduced the IMPACT Team Network approach, which combines a quality improvement approach with a leadership initiative to build supply chain leaders and skills, and introduces a structured framework for using data and analytics on a routine basis to drive decision making and action planning around supply chain improvement. We extracted all supply chain metrics possible from DHIS2 and analyzed and created actionable visualizations to facilitate easy interpretation and action by county managers to improve processes and outputs for the contraceptive and vaccine supply chains.

In partnership with the central Ministry of Health (MOH), JSI selected 10 counties that were representative of all geographies and contraceptive and vaccine coverage rates in the country and further conducted county engagement visits to determine interest, willingness, and commitment to adopting this approach.

## 3. RESULTS

JSI created 22 data dashboards in excel – two for each of the 10 counties and 2 for national level managers. Each county has one family planning and one vaccine dashboard, all of which have the same set of indicators, but which display data for that county only. The two national dashboards present data for the 10 counties alongside that for the 47 counties to allow for comparison of results over time. Dashboards were introduced to counties in June and July 2016 as part of the IMPACT Team rollout training. Both county dashboards are provided to the counties as part of the indicator tracking tool (ITT). County managers are being trained to use the data and the ITT for performance management, and to drive monthly analysis, problem solving and action planning. The ITT enables data extracted from DHIS2 to be organized, analyzed and visualized in a way that enables managers to easily understand the level of performance, identify the gap between current performance and desired or targeted performance. By integrating the IMPACT Team methodology into their existing meetings, county leaders are able to select a priority problem to focus on, conduct root cause analysis and identify practical actions for follow up against a timeframe, all aimed at improving the contraceptive and vaccine supply chain outcomes. The following month, once new data is in the ITT, the team reviews the same indicator to assess whether targeted performance levels have been achieved and if not, they repeat the process to identify different solutions or actions to address the same problem. If performance is satisfactory, they review other indicators to select a different priority and repeat the process. The methodology is encourages a culture of continuous improvement and the focus is on early wins to motivate the cycle to continue every month.

The Ministry of Health has both family planning and vaccine dashboards in their existing systems, although neither is able to display useful usable analytics and visuals for supply chain decision making. The MOH aims to incorporate the analytics and visuals from the ITT dashboards into its official dashboards, once counties have provided feedback and validated the utility of each metric and visual. Keeping in mind the existing dashboards, JSI deliberately created temporary dashboards in excel as an interim way in which to learn more about the utility of analytics from county managers, so as to avoid unnecessary duplication of tools and resources. The temporary excel dashboards are limited in that they must be manually updated each month with the previous month's data. Once the dashboards are incorporated into the online system, they will automatically be updated and immediately scaled to all counties in the country.

## 4. CONCLUSION

Effective use of data is vital to optimizing the performance of public health supply chains. As more and more countries automate health information and supply information management systems, a greater amount of data is available for use in achieving significant gains in efficiency and performance, if used effectively. However, effective date use requires managers to commit to governance structures that promote and enable transparency and accountability. The IMPACT Team Network approach marries these elements together towards the goal of improving public health supply chain performance and efficiency.

## 5. ACKNOWLEDGEMENTS

The author greatly appreciates the leadership and support of key colleagues from the Kenya Ministry of Health and the County Leadership and Commodity Security Technical Working Group Members from the 10 focus counties. The author also thanks colleagues from implementing partner organizations including the Clinton Health Access Initiative, IMA Health for their collaboration and partnership.